\documentclass[aps,pra,twocolumn,groupedaddress,showpacs]{revtex4}
\usepackage{color,amsmath, amssymb, graphicx, amsfonts , setspace}

\usepackage{subfigure}  
\newcommand{\abs}[1]{\ensuremath{\left| #1 \right|}}
\newcommand{\av}[1]{\ensuremath{\left\langle #1 \right\rangle}}

\newcommand{\mi}{\ensuremath{\mathrm{i}}} %to create an upright i for imaginary unit
\newcommand{\me}[1]{\ensuremath{\mathrm{e}^{#1}}} %upright e
\newcommand{\dif}{\ensuremath{\,\mathrm{d}}} %upright d for dx
\newcommand{\pdo}[2]{\ensuremath{\frac{\partial{#1}}{\partial {#2}}}} % pdo stands for partial derivative one (i.e. first partial derivative)
\usepackage{bm}
\renewcommand{\vec}[1]{\ensuremath{\bm{#1}}}%\mbox{\boldmath$#1$}}}

\newcommand{\nn}{\nonumber}

\newcommand{\dg}[0]{\ensuremath{\dagger}}

\newcommand{\ua}{\ensuremath{\uparrow}}
\newcommand{\da}{\ensuremath{\downarrow}}
\definecolor{refcolor}{rgb}{0.2,0.6,0.2}

\begin{document}

%\preprint{}

\title{Probing ultracold Fermi gases with light-induced gauge potentials}


\author{Jonathan M. Edge}
%\thanks{}
\affiliation{T.C.M. Group, Cavendish Laboratory, J.~J.~Thomson Ave., Cambridge CB3~0HE, UK.}
\affiliation{Instituut-Lorentz, P.O. Box 9506, NL-2300 RA Leiden, The Netherlands}
 \author{N. R. Cooper}
%\thanks{}
\affiliation{T.C.M. Group, Cavendish Laboratory, J.~J.~Thomson Ave., Cambridge CB3~0HE, UK.}

%\thanks{}
% \affiliation{}


\date{\today}

\begin{abstract}
 
  We theoretically investigate the response of a two component Fermi
  gas to vector potentials which couple separately to the two spin
  components. Such vector potentials may be implemented in ultracold
  atomic gases using optically dressed states. Our study indicates that
  light-induced gauge potentials may be used to probe the properies of
  the interacting ultracold Fermi gas, providing. amongst other
  things, ways to measure the superfluid density and the strength of
  pairing.
\end{abstract}

%67.85.-d, 03.75.Ss, 71.10.Pm, 73.20.Mf}
%\keywords{}

\maketitle


\section{Introduction}
\label{sec:introduction}

Experimental progress in the trapping and manipulation of ultracold
Fermi gases has led to new regimes of study of superfluid two
component Fermi systems. 
This is thanks to the unprecedented control over the microscopic properties of  many-body systems which ultracold Fermi gases offer. Examples of this microscopic control include the modification of the interaction strength via Feshbach resonances \cite{chin,bloch07}, the tuning of the density imbalance of  spin up and spin down particles \cite{fermionic_sf_with_imbalanced_populations-zwierlein-ketterle06,pairing_and_phase_sep_in_pol_fg_partridge_hulet} and the ability to impose controllable lattice potentials using optical lattices \cite{bloch07}. In particular the tunability of the interaction strength has allowed interesting investigations of the continuous crossover from a Bardeen-Cooper-Schrieffer (BCS) type superfluid, analogous to the BCS state in a superconductor, to a Bose Einstein condensate (BEC) of molecules consisting of bound spin up and spin down pairs \cite{leggett_orig_bec_bcs-xover1980,Observation_of_a_Strongly_Interacting_Degenerate_Fermi_Gas_of_Atoms-ohara-thomas-science02,prod_of_long_lived_ultracold_li_molecules-cubizolles03,Creation_of_ultracold_molecules_from_a_Fermi_gas_of_atoms-regal_nature03,pure_gas_of_opt_trapped_mol-jochim03,obs_of_bec_of_molecules_zwierlein03}.

Experiments on ultracold atomic Fermi gases allow studies of a variety of 
physical properties of a fundamental interest. Past work has studied the pairing gap as a function of the temperature and the interaction strength \cite{chin04}, the transition from superfluid to normal behaviour by studying collective modes \cite{Evidence_for_Superfluidity_in_a_Resonantly_Interacting_Fermi_Gas-kinast-thomas04} and the formation of vortices in rotating ultracold Fermi gases \cite{Vortices_and_superfluidity_in_a_strongly_interacting_FG-zwierlein-nature05}. 

Light-induced gauge potentials\cite{dalibardreview} offer the
possibility to study new aspects of the response of superfluid Fermi
system. This is the issue that we explore in this paper.  As we
describe below, light-induced vector potentials offer the opportunity
to study diverse properties of ultracold Fermi gases, ranging from the
superfluid density to properties of the pairs. These different
quantities become experimentally accessible by varying the time
dependence of the light induced vector potential and by considering
either the response to a vector potential which couples equally
(spin-symmetric) or in the opposite way (spin-asymmetric) to the two
spin species. Specifically, we decompose the vector
potentials $\vec A_{\ua\da}$ for the two spin species into a symmetric
component $\bar {\vec A}= \frac12 ( \vec A_\ua + \vec A_\da) $ and an
asymmetric component $\Delta \vec A = \vec A_\ua- \vec A_\da$.  We
study the response of a two-component Fermi gas, working in the BCS
limit, separately to the spin-symmetric $\bar {\vec A}$ and
spin-asymmetric $\Delta \vec A $ vector potentials.

The paper is organised as follows. In
Sec.~\ref{sec:effect-spin-symm} we describe how the spin-symmetric
vector potential $\bar {\vec A}$ leads to a superfluid response. In
Sec.~\ref{sec:effect-spin-asymm} we describe the response to a
spin-asymmetric vector potential $\Delta {\vec A}$ in the zero
temperature limit, exploring the frequency dependence in detail.  The
effects of non-zero temperature are described in
Sec.~\ref{sec:effects-non-zero}.  In Sec.~\ref{sec:exper-cons_vec_pot}
we explain how the effects we predict can be studied in
experiment. Finally, Sec.~\ref{sec:conclusion} summarizes our main
results.

\section{Effect of a spin-symmetric vector potential}
\label{sec:effect-spin-symm}

As discussed in \cite{measuring_sf_fraction_of_ultracold_atomic_gas-nigel_zoran10,john} for the bosonic system, the response of a one component Bose gas to a vector potential is determined by the superfluid density of the Bose gas. The light-induced vector potential leads to a kinetic energy of the form
$\frac{1}{2m}\left({\bm p} - {\bm A}\right)^2$,
which has its minimum shifted to ${\bm p} ={\bm A}$.
 The normal fluid component will seek the new minimum of the dispersion and thus come to rest at the new equilibrium. The superfluid component on the other hand is unaffected by the shifting of the minimum in the dispersion and will continue its initial state which now no longer is the equilibrium state. The phenomenology of a two component Fermi gas subjected to a spin-symmetric vector potential is the same as that of a one component Bose gas subjected to a vector potential.
For then the two components of the superfluid are affected equally by the vector potential and for the entire system there is a new steady state.
As in the Bose case, only the normal part of the superfluid will come to rest in the new steady state. The superfluid component will not relax to this new equilibrium state, but continue in its metastable state, thereby allowing one to distinguish between normal and superfluid densities.

As in the Bose case, the low-frequency response to $\bar {\vec A} $ is
simply determined by the superfluid density of the Fermi gas, with a
mean fermion number current density
\begin{equation}
\bar{\vec{j}} \equiv \frac12(\vec{j}_\ua + \vec{j}_\da) =  - \frac{\rho_s}{2m}\bar{\vec A}
\label{eq:sfdensity}
\end{equation}
where $\rho_s$ is the superfluid density, defined such that its maximum value is equal to the number density of fermions. 
We will therefore
concentrate on the response to $\Delta \vec A $. As we
shall show, 
in some regimes the response to $\Delta \vec A $ vanishes, under which conditions the response
can be determined from the response to
$\bar {\vec A} $ alone and is a measure of the superfluid density $\rho_s$.

\section{Effect of  a spin-asymmetric vector potential}
\label{sec:effect-spin-asymm}

\subsection{Static vector potential}
\label{sec:effect-static-vector}

We consider the BCS Hamiltonian for a system of spin $\frac12$ fermions with vector potentials $\vec A_\ua$ and $\vec A_\da$ coupling to the spin up and down particles respectively.
It is given by
\begin{align}
  H&=\int \dif^dr
  \left[
    \sum_\sigma
    \hat c^\dg_\sigma(\vec r) 
    \left(
      \frac
      {
        \left(
          {\hat{ \vec p} - \vec A_\sigma}
        \right)^2
      }{2m}
      -\mu
    \right)
    \hat c_\sigma(\vec r)
    \right.
    \nn \\ &
    \left.
    + V\hat c^\dg_\ua(\vec r)\hat c^\dg_\da(\vec r)\hat c_\da(\vec r)\hat c_\ua(\vec r)
  \right]
  \label{vec_pot:hamilt_in_real_space}
\end{align}
Here $c^{(\dg)}_{\sigma}(\vec r)$ are fermionic annihilation (creation) operators for spin $\sigma$ and
$V$ is the strength of the contact interaction $\hat V(\vec r-\vec
r')=V \delta(\vec r-\vec r')$. $V$ is related to the $s$-wave
scattering length $a_s$ via $V=\frac{4\pi\hbar^2a_s}m$ \cite{pitaevskii_stringari_RMP_09}. We will be considering attractive interactions, so $V<0$. For a uniform vector potential with $\vec A_\ua = -\vec A_\da =  \vec A $ the physical properties of the system are the same as for $\vec A=0$. In order to show this, 
perform the following transformation on the creation and annihilation operators.
\begin{equation}
    \hat c_\ua(\vec r) = \hat d_\ua(\vec r) \me{\mi \vec A_\ua \cdot \vec r} \quad\quad
    \hat c_\da(\vec r) = \hat d_\da(\vec r) \me{\mi \vec A_\da \cdot
      \vec r}
  \label{cha:vec_pot:ops_c_in_terms_of_d}
\end{equation}
Here and in the remainder  of this paper we set $\hbar=1$.
Upon inserting equations~(\ref{cha:vec_pot:ops_c_in_terms_of_d}) into equation~(\ref{vec_pot:hamilt_in_real_space}) we recover the same Hamiltonian but now for the operators $\hat d_\sigma $
\begin{align}
  H&=\int \dif^dr
  \left[
    \sum_\sigma
    \hat d^\dg_\sigma(\vec r) %\epsilon_{k,\sigma}
    \left(
      \frac{\hat{\vec p}^2}{2m} -\mu
    \right) \hat d_\sigma(\vec r) 
   \right.
    \nn \\ &
    \left.
    + V\hat d^\dg_\ua(\vec r)\hat d^\dg_\da(\vec r)\hat d_\da(\vec r)\hat d_\ua(\vec r)
  \right]
  \label{vec_pot:hamilt_in_real_space_in_terms_of_d}
  %\int\dif^d\vec r 
\end{align}
For $\vec A_\ua= -\vec  A_\da=\vec A$ the mean-field ground state quantities $\rho_A(x) $ and $\Delta_A(x) $ are the same as the quantities $\rho$ and $\Delta$ for $A_\sigma = 0$. The densities are given by
\begin{align}
  \rho_A(\vec r) &= \av{\hat c^\dg_\sigma(\vec r) \hat c_\sigma(\vec r)} = \av{\hat d^\dg_\sigma(\vec r) \hat d_\sigma(\vec r) \me{\mi (\vec A_\sigma -\vec  A_\sigma) \cdot \vec r}}
  \nn\\
  &= \rho_0(\vec r)
  \label{rho_A_in_terms-of_rho}
\end{align}
and the gap is given by \cite{pitaevskii_stringari_RMP_09}
\begin{align}
  \Delta_A(\vec r) &= V_{\text{eff}}\av{\hat c_\da(\vec r) \hat c_\ua(\vec r)}
  = V_{\text{eff}}\av{\hat d_\da(\vec r) \hat d_\ua(\vec r)
    \me{\mi (\vec A_\da +\vec  A_\ua) \cdot \vec  r}}
\nn\\
  =& \Delta(\vec r) = \Delta_0
  \label{Delta_A_in_terms-of_A}
\end{align}
since $\Delta$ is homogeneous. 
We have introduced an effective potential $V_{\text{eff}}$ to regularise the divergent nature of the contact interaction \cite{pitaevskii_stringari_RMP_09}.
This shows that for $\vec A_\ua= - \vec A_\da= \vec A $ the ground state properties are indeed independent of $\vec A $.

This is to be contrasted with the result for $\vec A_\ua= \vec A_\da$. In that case the ground state of the system is a state in which the phase of the gap is spatially varying.
This can be seen from equation~(\ref{Delta_A_in_terms-of_A}). If we insert $\vec A_\ua= \vec A_\da= \vec A $ into equation~(\ref{Delta_A_in_terms-of_A}) we obtain
\begin{align}
  \Delta_A(\vec r) &= V_{\text{eff}}\av{\hat c_\da(\vec r) \hat c_\ua(\vec r)}
  = V_{\text{eff}}\av{\hat d_\da(\vec r) \hat d_\ua(\vec r)
    \me{\mi (\vec A_\da +\vec  A_\ua) \cdot \vec  r}}
  \nn\\
  &= \Delta_0\, \me{{2\mi} \vec A\cdot\vec r}
  .
  \label{Delta_A_in_terms-of_A_A_up_eq_A_down}
\end{align}
showing that the phase of the gap is $\varphi = {2\mi} \vec A\cdot\vec r$ and therefore spatially varying
for nonzero ${\bm A}$. Although the gap is spatially varying in the groundstate, the {\it gauge-invariant} supercurrent density remains zero.
The density also remains unchanged, as can be seen from equation~(\ref{rho_A_in_terms-of_rho}).

\subsection{Time-dependent vector potential}
\label{sec:effect-time-depend}

Given that the application of a static vector potential $\vec A $ with
$\vec A_\ua = -\vec A_\da =  \vec A $ does not have an effect on the
ground state of the system we consider a time-dependent vector
potential $\vec A $. In the analogous electromagnetic system
$\pdo{\vec A}t \not=0 $ corresponds to the presence of an electric
field \cite{jackson_2nd_ed}. However, since we are considering $\vec A_\ua= - \vec A_\da$ this would correspond to a Fermi system where the spin up and spin down particles have opposite charges. An example would be a spin polarised electron-positron superfluid where the spin up particles are positrons and the spin down particles are electrons.

We now treat the application of time-dependent vector potentials $\vec A_{\sigma}(t) $ within linear response. Write the Hamiltonian as
\begin{align}
  \hat H = \hat H_0 + \delta \hat H(t) .
\end{align}
$H_0$ is now the mean-field BCS Hamiltonian in the absence of a vector potential which we consider in second quantised form in momentum space. It is given by 
\begin{align}
  H_0 &= \sum_{\vec k,\sigma}
  \left(
    \frac{k^2}{2m} - \mu
  \right) \hat c^\dg_{\vec k,\sigma} \hat c_{\vec k,\sigma}
  +\sum_{\vec k}
  \left(
    \Delta \hat c^\dg_{\vec k,\ua}\hat c^\dg_{- \vec k,\da} + h.c.
  \right).
\end{align}
$\Delta $ is determined self-consistently by the equation~\cite{pitaevskii_stringari_RMP_09}
\begin{align}
  \frac1{V} &= \frac1{2L^d} \sum_{\vec k}
  \left(
    \frac1{E_k} - \frac{2m}{k^2}
  \right)
\end{align}
where $L$ is the size of the system. 
To linear order in $\vec A$, $\delta \hat H $ is given by
\begin{align}
  \delta \hat H(t) &= - \sum_{\vec k}
  \left(
    \frac{\vec k \cdot \vec A_\ua(t)}{m} \hat c^\dg_{\vec k,\ua}(t) \hat c_{\vec k,\ua}(t)
  \right.
  \nn\\* &\qquad+
  \left.
 %   +
    \frac{\vec k \cdot \vec A_\da(t)}{m} \hat c^\dg_{\vec k,\da}(t) \hat c_{\vec k,\da} (t)
  \right)
  .
  \label{vec_pot:delta_H}
\end{align}
The response of an observable $\hat M$ is then given by the Kubo formula \cite{forster75}
\begin{align}
  \delta \av {\hat{M} (t)} =  \frac 1\mi \int_{-\infty}^t \dif\tau \av{
    \left[
      \hat M(t), \delta \hat H (\tau)
    \right]
  }
  .
  \label{kubo_formula}
\end{align}
We consider $\vec A_\sigma(t)= \vec A_\sigma\me{\mi\omega t} $ as the time-dependent perturbation,
and study the current response.
As we are applying a spin-asymmetric vector potential we look at the spin current $\Delta\!\hat{\vec J}=\hat{\vec J}_\ua-\hat{\vec  J}_\da$. Linear response of the current involves two contributions. 
First, there is the contribution from the Kubo formula, equation (\ref{kubo_formula}), with $\hat{M}$ replaced by the spin current operator in the absence of  a vector potential  $\vec A$ 
\begin{align}
  \Delta \hat {\vec J}(t) &=
  \sum_{\vec k}\frac{\vec k }{m}
  \left(
    \hat c^\dg_{\vec k,\ua}(t) \hat c_{\vec k,\ua}(t) - \hat c^\dg_{\vec k,\da}(t) \hat c_{\vec k,\da} (t)
  \right)
  .
  \label{def_of_delta_J}
\end{align}
Secondly, there is a contribution arising from the redefinition of the spin current for nonzero $\vec A$.
 Applying the vector potential $\vec A $ changes the current operator to the gauge invariant current operator
\begin{align}
  \hat{\vec J}_\sigma=\sum_{\vec k}\frac{\vec k - \vec A_\sigma}{m}\hat c^\dg_{\vec k,\sigma}(t) \hat c_{\vec k,\sigma}(t)
  .
\end{align}
This means that
definition of the current changes such that
\begin{align}
  \hat{\vec J}_\sigma \to\hat{\vec J}_\sigma- \sum_{\vec k} \frac{\vec A_\sigma}{m} \hat c^\dg_{\vec k,\sigma} \hat c_{\vec k,\sigma}
  .
  \label{redef_of_current}
\end{align}
Inserting equations~(\ref{vec_pot:delta_H}),~(\ref{def_of_delta_J}) and (\ref{redef_of_current}) into equation~(\ref{kubo_formula}) we obtain
for the spin current in linear response
\begin{widetext}
\begin{align}
  \av{\Delta\!\vec J} (t)&=
  - \sum_{\vec k} \frac{\vec A_\ua}{m} \av{\hat c^\dg_{\vec k,\ua} \hat c_{\vec k,\ua}}
  + \sum_{\vec k} \frac{\vec A_\da}{m} \av{\hat c^\dg_{\vec k,\da} \hat c_{\vec k,\da}}
  + \frac 1\mi \int_{-\infty}^t \dif\tau
  \me{-\eta(t-\tau)}
  \av{
    \left[
      \sum_{\vec k}\frac{\vec k }{m}
      \left(
        \hat c^\dg_{\vec k,\ua}(t) \hat c_{\vec k,\ua}(t) - \hat c^\dg_{\vec k,\da}(t) \hat c_{\vec k,\da} (t)
      \right),
    \right.
  \right.
  \nn\\* &\quad\quad
  \left.
    \left.
      + \sum_{\vec k'}
      \left(
        \frac{\vec k' \cdot \vec A_\ua(t)}{m} \hat c^\dg_{\vec k',\ua}(\tau) \hat c_{\vec k',\ua}(\tau) +
        \frac{\vec k' \cdot \vec A_\da(t)}{m} \hat c^\dg_{\vec k',\da}(\tau) \hat c_{\vec k',\da} (\tau)
      \right)
    \right]
  }
  .
  \label{first_expr_J_up_-J_down}
\end{align}
\end{widetext}
The term $\me {-\eta (t -\tau)} $ serves to regularise the expression and physically means that the perturbation is gradually switched on starting at $T = - \infty$.
We will eventually take the limit $\eta\to 0 $.

To evaluate the expressions in equation~(\ref{first_expr_J_up_-J_down}) we expand the operators $\hat c_{\vec k,\sigma} $ in terms of the quasiparticle operators $\hat \alpha_{\vec k,\sigma} $ via \cite{ben_and_atland06}
\begin{align}
  \begin{pmatrix}
    \hat c_{\vec k,\ua}\\
    \hat c^\dg_{-\vec k,\da}
  \end{pmatrix}
  =
  \begin{pmatrix}
    u_k & v_k\\
    v_k &-u_k
  \end{pmatrix}
  \begin{pmatrix}
    \hat \alpha_{\vec k,\ua}\\
    \hat \alpha^\dg_{-\vec k,\da}
  \end{pmatrix}
\end{align}
where $u_k=  \sqrt{\frac{1 + \frac{\epsilon_k}{E_k}}2}$ and $v_k=  \sqrt{\frac{1 - \frac{\epsilon_k}{E_k}}2}$ with $\epsilon_k = \frac{k^2}{2m} - \mu$ and $E_k = \sqrt{\epsilon_k^2+ \Delta^2}$.
The time dependence of the operators $\hat \alpha_{\vec k,\sigma} $ is given by $\hat \alpha_{\vec k,\sigma}^\dagger(t) = \hat \alpha_{\vec k,\sigma}^\dagger \me{\mi E_k t}$ and $\hat \alpha_{\vec k,\sigma} (t)= \hat \alpha_{\vec k,\sigma} \me{-\mi E_k t}$ \cite{deGennes66}.
Inserting these gives  the response
\begin{align}
  \av{\Delta\!\vec J}(t)&=  \frac 2{\mi m^2} \int_{-\infty}^t \sum_{\vec k}
  - \vec k \me{-\eta(t-\tau)}
  u_k^2 v_k^2 
  \left(
    2f(E_k) -1
  \right)
   \nn\\ &
  \left\{
    \me{2\mi E_k (t-\tau)} - \me{-2\mi E_k (t-\tau)}
  \right\}
  \nn\\* &
  \left(
    \vec k \cdot \vec A_\ua(\tau) -\vec k\cdot \vec A_\da(\tau)
  \right)
   \nn\\ &
  - \sum_{\vec k}
  \left(
    v_k^2 + f(E_k) \frac{\epsilon_k}{E_k}
  \right)
  \frac1m
  \left(
    \vec A_\ua(t)-\vec A_\da(t)
  \right)
 ,
  \label{delta_J_in_time_domain}
\end{align}
where $f$ is the Fermi function.
Taking $\vec A_\ua - \vec A_\da$ to point along the $z$-axis and converting the sum to an integral via $\sum_{\vec k} \to
\left(
  \frac{L}{2\pi}
\right)^d\int\dif^dk$ 
and Fourier transforming gives for $d = 3 $
\begin{align}
  \av{ \Delta\!J_z}(\omega)&=
  \left\{
    \frac2{m^2}
    \left(
      \frac{L}{2\pi}
    \right)^3
  \right.
  \nn\\ &
  \left.
    \frac{4\pi}3
    \int_0^\infty\dif k k^4 
    \frac{\Delta^2}{4E_k^2}
    (1-2f(E_k))
  \right.
  \nn\\ &
  \left.
    \left(
      \frac1{2E_k + \omega + \mi \eta }+  \frac1{2E_k - \omega - \mi \eta  }
    \right) 
  \right.
  \nn\\ &
  \left.
    - \frac1m 
    \left(
      \frac{L}{2\pi}
    \right)^3
    4\pi \int_0^\infty k^2 \dif k
    \left(
      v_k^2 + f(E_k) \frac{\epsilon_k}{E_k}
    \right)
  \right\}
   \nn\\ &
  \left(
    {A_\ua(\omega) - A_\da(\omega)}
  \right)
  \label{resp_of_J_in_freq_space1}
  \\
  &= \chi (\omega) \Delta\! A(\omega)
\end{align}
where $\chi (\omega) $ is the response function for $\Delta J$ along the  $z$-direction. In the zero temperature limit $\chi (\omega) $ can  be expressed as
\begin{align}
  \chi (\omega) &=  \frac1{m^2}
  \left(
    \frac{L}{2\pi}
  \right)^3
  \frac{2\pi}3
  \left(
    2m
  \right)^{5/2} \Delta^{3/2}
  \nn\\* &
  \int_0^\infty \dif x \frac {x^4}{(x^2 - a^2)^2 + 1}
  \left(
    \frac1{2\sqrt{(x^2 - a^2)^2 + 1} +\tilde\omega+ \mi\eta} 
    \right.
    \nn\\ &\qquad\qquad
    \left.
    + \frac1{2\sqrt{(x^2 - a^2)^2 + 1} -\tilde\omega - \mi\eta}
  \right)
  \nn\\* &%\qquad
  - \frac1m 
  \left(
    \frac{L}{2\pi}
  \right)^3
  2\pi
  \left(
    2m\Delta
  \right)^{\frac32}
  \nn\\* &\qquad
  \int_0^\infty x^2 \dif x
  \left(
    1-\frac{x^2-a^2}{\sqrt{
        \left(
          x^2-a^2
        \right)^2 +1
      }
    }
  \right)
  \label{resp_integral_dimensionless}
\end{align}
where $a ^ 2 = \frac\mu\Delta$ and $\tilde\omega = \frac\omega\Delta$. In the limit $\tilde\omega\ll 1$ we can find an analytic expression for $\chi(\omega)$. First expand the integral in equation~(\ref{resp_integral_dimensionless})  for $\tilde\omega\ll 1 $ 
\begin{align}
  &\int_0^\infty \dif x \frac {x^4}{(x^2 - a^2)^2 + 1}
%   \nn\\ &
  \left(
    \frac1{2\sqrt{(x^2 - a^2)^2 + 1} +\tilde\omega + \mi\eta}
  \right.
  \nn\\&\qquad
  \left.
    + \frac1{2\sqrt{(x^2 - a^2)^2 + 1} -\tilde\omega - \mi\eta}
  \right)
  \nn\\
  &= \int_0^\infty \dif x \frac {x^4}{((x^2 - a^2)^2 + 1)^{3/2}}
  \nn\\ &\qquad
  + \frac{\tilde\omega^2}{4} \int_0^\infty \dif x \frac {x^4}{((x^2 - a^2)^2 + 1)^{5/2}}
  + O(\omega^4)
  .
  \label{vec_pot:expanding_integral}
\end{align}
Now note that for $T=0$ the $\omega ^ 0 $ part of this integral cancels the second term in eq.~(\ref{resp_integral_dimensionless}) 
coming from the redefinition of the current. The integral in the second term in eq.~(\ref{vec_pot:expanding_integral}) can be obtained from \cite{gradshteyn_ryzhik_tables_of_integrals_series_and_products}
$
   \int_0^\infty\frac {x^4 \dif x}{ (x^4 + 2b^2 x^2 + c^4)^{3/2}} = K(d) \frac{c}{2(c^2-b^2)} - E(d) \frac{cb^2}{c^4-b^4}
$
by parametric differentiation with respect to $c $. Here $d$ is given by $d =\frac{\sqrt{c^2-b^2}}{\sqrt2 c} $ and $K (x) $ and $E (x) $ are complete elliptic integrals of the first and second kind and are defined as
\begin{align}
 K (x) &=\int_0^{\frac\pi2}\frac{1}{\sqrt{1-x^2\sin^2\phi}}\dif \phi\\
 E (x) &= \int_0^{\frac\pi2}{\sqrt{1-x^2\sin^2\phi}}\dif \phi .
\end{align}
Defining $r=
\frac{\sqrt{\left(a^4+1\right)^{1/2} + a^2}}{\sqrt2 \left(a^4+1\right)^{1/4}}$ we 
finally  obtain the following for the zero temperature response function in the limit $\frac\omega\Delta\ll 1$
\begin{widetext}
  \begin{align}
    \chi (\omega) &=\frac1{m^2} \left( \frac{L}{2\pi} \right)^3
    \frac{2\pi}3 \left( 2m \right)^{5/2} \Delta^{3/2}
    \frac{\tilde\omega^2}{24 (a^4+1)^{3/4}} 
    \nn\\& 
    \left[ \frac12 K(r)
      \left( (a^4 + 1)^{3/2} + 3a^4(a^4+1)^{1/2} - 4a^2 (a^4+1)
      \right)
      % \right.  \nn\\* & \left.
      + 4 E(r) (a^4 +1)a^2 \right] .
    \label{vec_pot:chi_T_0_omega_small}
  \end{align}
\end{widetext}
As the BCS approximation we are working in is only valid up to order $\frac1{a^2}$ we expand equation~(\ref{vec_pot:chi_T_0_omega_small}) to lowest order in $\frac1{a}$ and obtain 
\begin{align}
  \chi (\omega) &=  \left( \frac{L}{2\pi} \right)^3
    \frac{4\pi\sqrt{2m}}9  \mu^{3/2}
    {\tilde\omega^2}
    .
    \label{expansion_for_chi_large_a}
\end{align}

Equation~(\ref{vec_pot:chi_T_0_omega_small}) shows that in the limit $\omega\to 0 $, \mbox{$\chi (\omega, T=0) $} vanishes quadratically with $\omega $. We thus find that the low-frequency response is consistent with the result in section~\ref{sec:effect-static-vector}, namely that a static vector potential acting in the opposite way on the two spin species has no effect on the ground state of the system.

\subsection{Interpretation of the low-frequency response}

The low-frequency response of the system can be understood in terms of currents arising from the time-varying polarisation of the Cooper pairs in the Fermi gas.  A time dependent vector potential $\vec A $ gives rise to an effective electric field $\vec E $ via $\vec E = - \pdo{\vec A} t $. Since we are considering $\vec A_\ua = - \vec A_\da$ this gives rise to an electric field which couples differently to the spin up and spin down particles, such that we obtain
\begin{align}
  \Delta\!\vec E = - (\dot{\vec A}_\ua-  \dot{\vec A}_\da) = -\mi\omega\Delta\! \vec A
  .
  \label{E_as_time_deriv_of_A}
\end{align}
In the absence of such an effective electric field the s-wave Cooper pairs which make up the superfluid are spherically symmetric. Application of an effective electric field leads to a displacement of the average positions of the spin up and spin down particles. This leads to a polarisation of the Cooper pairs just like the application of an electric field to a hydrogen atom induces a dipole moment and hence a polarisation in the hydrogen atom. A time-dependent polarisation results in a current which is given by \cite{jackson_2nd_ed}
\begin{align}
  \vec J =\pdo {\vec P}t
  .
  \label{J_as_time_deriv_of_Polarisation}
\end{align}
Combining equation~(\ref{E_as_time_deriv_of_A}) and equation~(\ref{J_as_time_deriv_of_Polarisation}) we obtain  the polarisation
\begin{align}
  P =\frac\chi {\omega ^ 2}\Delta E
  ,
  \label{polarisability}
\end{align}
which implies that $\frac\chi {\omega ^ 2} $ is the polarisability $\gamma $. According to equation~(\ref{expansion_for_chi_large_a}) $\gamma $ is constant for small $\omega $ and proportional to $\Delta ^ {-2} $.
The polarisability describes how easily the particles in the
spherically symmetric s-wave Cooper pairs can be displaced with
respect to each other. A larger gap $\Delta $ means that the Cooper
pairs are more tightly bound and the system has a reduced
polarisability.

Note that the polarisability we describe here is very different from
the one described in Ref. 
\onlinecite{dipole_pol_of_trapped_sf_FG-recati_stringari06}. In this paper
the polarisability we are considering is the dynamic polarisability
arising from the individual Cooper pairs in the absence of pair
breaking effects. In
Ref. \onlinecite{dipole_pol_of_trapped_sf_FG-recati_stringari06} on
the other hand the static polarisation of an atomic Fermi gas cloud is
considered which arises when the two spin species are subjected to
different potentials. The polarisation predicted in Ref. \onlinecite{dipole_pol_of_trapped_sf_FG-recati_stringari06} occurs only once Cooper pairs have been broken.

\subsection{Beyond the low-frequency limit}
\label{sec:beyond-low-frequency}

The above analysis was done for $\frac\omega\Delta\ll1 $. For larger values of $\omega $ we need to solve equation~(\ref{resp_integral_dimensionless}) numerically.

\begin{figure}[htb]
  \centering
  \includegraphics[width=8cm]{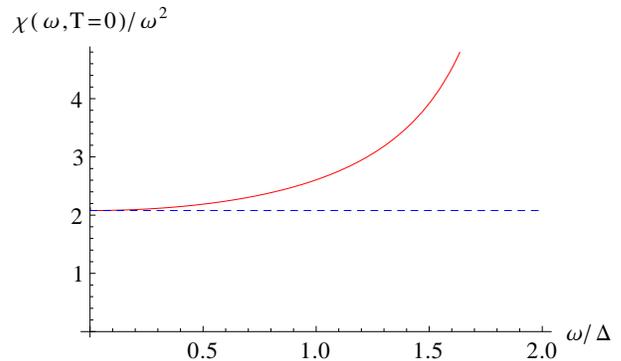}
  \caption{(Colour online) Response function $\chi$ divided by $\omega^2$ and normalised to the number of particles as a function of $\frac\omega\Delta$ in the region $0\le\omega<2\Delta$. The interaction strength is given by $k_Fa_s=-1.1$ where $a_s=\frac{mV}{4\pi\hbar^2}$ is the $s$-wave scattering length. The parameter  $a=\frac \mu\Delta$ is given by $a=3.7$ and $r
%\frac{\sqrt{\left(a^4+1\right)^{1/2} + a^2}}{\sqrt2 \left(a^4+1\right)^{1/4}} 
=1.0$.
  Solid Line: $\frac{\chi(\omega)}{\omega^2}$ as obtained by numerically integrating the expression in equation~(\ref{resp_integral_dimensionless}). Dashed line: Expression obtained from equation~(\ref{vec_pot:chi_T_0_omega_small}).}
  \label{fig:resp_div_by_omega_sq}
\end{figure}
\begin{figure}[htb]
  \centering
  \includegraphics[width=8cm]{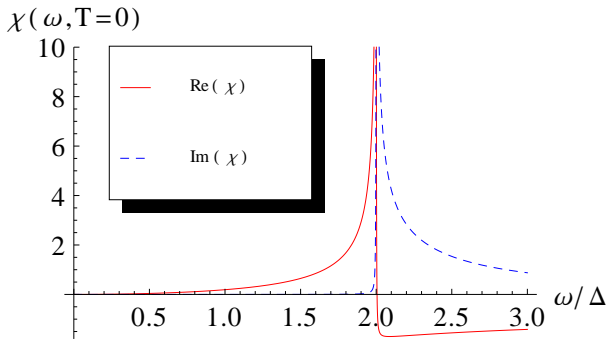}
  \caption{(Colour online) Real and imaginary part of the response function for $T=0$ as a  function of $\frac\omega\Delta$ for the same system as in fig.~\ref{fig:resp_div_by_omega_sq}. In the limit $\eta\to0$ there is an infinitely sharp peak at $\omega=2\Delta$. Above $\omega=2\Delta$ the response  is independent of $\omega$ and is given by equation~(\ref{redef_of_current}).}
  \label{fig:resp_for_0_sm_omega_smaller_3Delta}
\end{figure}

As can be seen from equation~(\ref{resp_integral_dimensionless}), for $\omega < 2\Delta $ the response is purely real in the limit $\eta\to 0 $. Since the imaginary part of the response function describes the energy absorbed by the system \cite{nozieres_Fermi_liquids2} this means that no energy is absorbed. This is what is  expected since the perturbation does not have enough energy to break pairs.
We have plotted the response function divided by $\omega ^ 2 $ in the range $0 \leq\omega < 2\Delta $ in figure~\ref{fig:resp_div_by_omega_sq} for a particular set of parameters. This shows that for a reasonably wide range of $\omega $ the response is indeed well described by equation~(\ref{vec_pot:chi_T_0_omega_small}). Significant departures occur only for $\frac\omega\Delta \gtrsim 0.5 $.
In figure~\ref{fig:resp_for_0_sm_omega_smaller_3Delta}   the real and imaginary parts of $\chi(\omega) $ for $0 <\omega < 3\Delta  $ are plotted.
As $\omega\to 2\Delta $ the real part of $\chi $ diverges  as $(2\Delta-\omega)^{-1/2}$. At $\omega = 2\Delta $ the real and imaginary parts of $\chi $ diverge with $\eta$ as $\eta ^ {-1/2} $. The energy absorption of the system diverges since the perturbation can resonantly couple to pair breaking processes. For $\omega > 2\Delta $ the perturbation has enough energy to break pairs but is no longer on resonance. The energy absorption in this regime is proportional to $(\omega^2-4\Delta^2)^{-1/2} $.

The real part of the response for $\omega > 2\Delta$ is dominated by the contribution from the redefinition of the current, see equation~(\ref{redef_of_current}). This is because the perturbation has enough energy to break pairs and once pairs are broken they behave like  non-interacting particles, which, in the absence of collisions, will not equilibrate to the new lowest energy state. This leads to response 
\begin{equation}
\Delta {\bm J} = -\frac{N}{2m}\Delta {\bm A}
\label{eq:barecurrent}
\end{equation}
where $N$ is the total number of atoms. This behaviour of non-interacting particles can be seen from equation~(\ref{resp_of_J_in_freq_space1}) by setting $\Delta = 0 $.

\section{Effects of non-zero temperature}
\label{sec:effects-non-zero}

\subsection{Collisionless regime}
\label{sec:collisionless-regime}

\begin{figure}[htb]
  \centering
   \includegraphics[width=8cm]{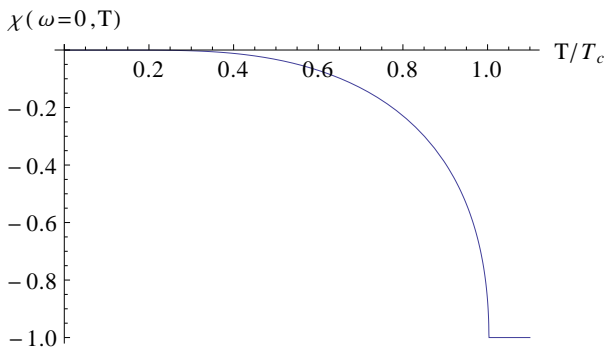}
  \caption{(Colour online) Response as a function of $\frac T{T_c}$ both below and above $T_c$ for the same system as in fig.~\ref{fig:resp_div_by_omega_sq}. Normalisation is with respect to the response of a noninteracting gas, $N/2m$. The kink in the response function occurs at $T=T_c$.}
  \label{fig:vec_pot_resp_as_fn_of_T}
\end{figure}

For non-zero temperature there is a current response even at zero frequency. This can be understood by noting that at non-zero temperature there are thermally excited quasiparticles and within the approximation we are working in thermally excited quasiparticles do not interact. The important effects of quasiparticle collisions will be treated in section~\ref{sec:quas-scatt-rate}.
For non-interacting quasiparticles the only contribution to the current comes from the redefinition of the current. Within this approximation, the response function then measures the {\it density} of unpaired quasiparticles, reaching its maximum magnitude at $T\geq T_{\rm c}$ when all particles are unpaired.
For temperatures $\frac T{T_c} \ll 1 $ where $T_c $ is the critical temperature $\chi(\omega\to0) \sim \exp{\left(-\frac\Delta T\right)} $. On the other hand, close to $T_c $ we find that $\chi(\omega\to 0,T) -\chi(\omega\to 0,T_c)  \sim  \abs{T-T_c}^\beta$ where within mean-field theory $\beta $, the critical exponent for $\Delta $,  is given by $\beta = 1/2 $ \cite{ben_and_atland06}. The response as a function of temperature  at $\omega = 0 $ is shown in figure~\ref{fig:vec_pot_resp_as_fn_of_T}.
Above $T_c $ we have $\Delta = 0 $ and hence the whole gas behaves like a gas of non-interacting particles whose response is simply  given by the redefinition of the current, equation~(\ref{redef_of_current}), for the full density leading to (\ref{eq:barecurrent}).

\subsection{Quasiparticle collisions}
\label{sec:quas-scatt-rate}

In the above derivation of the response we have neglected quasiparticle collisions.
At non-zero temperature thermally excited quasiparticles are present in the Fermi gas. A thermally excited quasiparticle with momentum $\vec p_1 $ and energy $\epsilon_1 -\mu  $ can undergo a scattering process
\begin{align}
  \vec p_1 + \vec p_2 \to \vec p_1' + \vec p_2'
\end{align}
where the other particle with momentum $\vec p_2$ can be situated above or below the Fermi surface.
This leads to a non-zero collisional relaxation rate $\Gamma $. As the temperature tends to zero the lifetime of an excited quasiparticle diverges due to Pauli blocking.
Unless $\omega \gg\Gamma $ we cannot neglect quasiparticle scattering.

In order to obtain a rough estimate for the collisional relaxation rate $\Gamma $,
we consider the quasiparticle scattering rate in the Fermi liquid picture
\cite{theory_of_quantum_liquids1_nozieres_1966}. 
This is an appropriate estimate for $T \sim T_{\rm c} \ll T_F$ when a sizeable density of unpaired quasiparticles is thermally excited.
Within perturbation theory the quasiparticle scattering  rate  for $T
\ll T_F $ for a thermally excited quasiparticle  with energy $\epsilon
-\mu = k_BT $ and for a $\delta $ function interaction potential $\hat
V = V\delta ^ 3 (\vec r - \vec r') $,
cf. equation~(\ref{vec_pot:hamilt_in_real_space}), can be derived
along the lines of Ref. \onlinecite{theory_of_quantum_liquids1_nozieres_1966}. We find
\begin{align}
  \Gamma= \mu
  \left(
    k_F a_s
  \right)^2
  \left(
    \frac T {T_F}
  \right)^2
  \frac{\pi^2+1}\pi  \frac{\me{}}{\me{} +1}.
 \label{eq:gamma}
\end{align}
For $T\ll T_{\rm c}$ the quasiparticle density is suppressed by the superfluid pairing, so the scattering rate is reduced below this estimate.

The results in figure~\ref{fig:vec_pot_resp_as_fn_of_T} are accurate whenever $\Gamma \ll\omega $ and $\omega \ll\Delta $. In the BCS regime of weak coupling, with $T_{\rm c}\simeq \Delta \ll \mu$, these conditions can both be satisfied for $T\sim T_{\rm c}$ where the estimate  (\ref{eq:gamma}) applies.

When $\omega \ll\Gamma $,  notably for the limit $\omega \to 0$ of a static perturbation, the quasiparticle collisions serve to damp the counterpropagating currents $\Delta J = J_\ua- J_\da$. This causes the zero-frequency response to vanish for all temperatures. At small non-zero frequency, this damping will lead to a dissipative spin current, with $\Delta J \propto  i \omega \Delta A / \Gamma$.

The fact that for $\omega\to 0 $ the response is zero even for $T > 0 $ in the presence of quasiparticle scattering is very important if one wishes to measure the superfluid density in the way outlined in section~\ref{sec:effect-spin-symm}.
For that method to work one requires the response of $\Delta\vec J $ to a perturbation $\Delta A $ to be zero. Figure~\ref{fig:vec_pot_resp_as_fn_of_T} indicates that in the absence of quasiparticle collisions $\chi (\omega,T) $ becomes appreciable, a few percent of the maximal (non-interacting) value, for $\frac T {T_c} \gtrsim \frac13 $. On the other hand, as long as we can assume that $\Gamma $ is much larger than $\omega $, quasiparticle collisions will serve to ensure that $ \Delta \vec J=0$. Thus if a vector potential is applied to only one species, the spin current will be zero and the total current can be used to infer the superfluid density.

This suppression of $\Delta J $ at low temperature can be understood
as a form of spin drag. A vector potential $A_\ua$ acting on the spin
up particles causes a current $J_\ua $ which in turn causes a current
$J_\da $, eventually (at low enough frequency) leading to a
suppression of $\Delta J $. In non-superfluid ferromagnetic Fermi
gases the spin drag has been investigated in Ref. \onlinecite{duine_viginale_spin_drag_FG2010}. The main difference in the superfluid systems we are considering here is that at $\omega = 0 $ there is no relaxation of the spin drag. This is because unless Cooper pairs are broken at $\omega\not = 0 $, the constituents of the Cooper pairs cannot move independently.

\section{Experimental considerations}
\label{sec:exper-cons_vec_pot}

The most natural experimental implementation of these ideas involves a
modification of the geometry described in
Ref. \onlinecite{measuring_sf_fraction_of_ultracold_atomic_gas-nigel_zoran10}
to the two-species Fermi gas. Thus, we consider the two species
confined to a ring-like trap. The application of coherent optical
fields, using beams of non-zero angular
momentum\cite{measuring_sf_fraction_of_ultracold_atomic_gas-nigel_zoran10},
allows azimuthal vector potentials to be imprinted on one, or both, of
these species. Clearly the simplest case is to dress just one
species (say spin-$\ua$), giving $\vec{A}_\ua\neq0$ and $\vec{A}_\da
= 0$.  The methods described in
Ref. \onlinecite{measuring_sf_fraction_of_ultracold_atomic_gas-nigel_zoran10},
then allow the average azimuthal momentum of the dressed species,
and hence, its current $J_\ua$, to be measured spectroscopically.
Temporal modulation of the amplitude of these optical fields and/or
the detuning allows $\vec{A}_\ua$ to be time-dependent, allowing the
use of this technique to probe the response at nonzero frequency.

In order to understand the consequences of the results described
above, we express the vector potentials in terms of the symmetric
component $\bar{\vec A} =\frac12 (\vec A_\ua+ \vec A_\da) $ and a spin
asymmetric component $\Delta \vec A = (\vec A_\ua- \vec A_\da) $. When
just one species is dressed, $\vec{A}_\ua \equiv \vec{A}$ and
$\vec{A}_\da = 0$, these components are both non-zero $\bar{\vec A} =
\frac12 {\vec A}$ and $\Delta \vec A = \vec A$.

We have shown that in the low frequency limit $\omega\ll \Gamma$ 
the response to $\Delta {\vec A}$ is zero for any temperature. 
This implies that the system only
shows a current response to $\bar {\vec A} $. This induces a linear response of the superfluid, with a current 
density (\ref{eq:sfdensity}) set by the superfluid density $\rho_s$.
The non-equilibrium
state of the superfluid causing this steady state flow remains (meta)stable provided the induced velocity remains below the superfluid critical velocity.
For optical dressing of a single species, $\bar{\vec A} = \frac12
{\vec A}$ and $\Delta \vec A = \vec A$, the resulting current density
of this species is ${\vec j}_\ua = \bar{\vec{ j}} + \frac12\Delta {\vec j} =
-\frac{\rho_s}{2m}\bar{\vec A} = -\frac{\rho_s}{4m} {\vec A}$. 
Thus, a spectroscopic measurement of
the total current  of this species ${\vec J}_\ua = {\cal V} {\vec j}_\ua$, where ${\cal V}$ is the
total volume, allows a direct measurement of the superfluid
density $\rho_s$.  The superfluid density of a harmonically trapped
Fermi gas has been measured using the collective
modes \cite{grimm_sf_density_2009}. The method we propose here can be
applied in a wider range of geometries (it does not rely on the
harmonicity of the trap). With local imaging, it would also be able to
probe the local superfluid density in different parts of the atom
cloud. Furthermore, the method allows interesting additional
information to be obtained from the response at non-zero frequencies.

For $\omega \gtrsim \Gamma$ the response to $\Delta\vec A $ is
non-zero and becomes appreciable for $\omega \simeq \Delta$, as shown
in Fig. \ref{fig:vec_pot_resp_as_fn_of_T}.  For $\omega <2\Delta $ but
not $\omega \ll \Delta $ the response of the spin current $J_\ua-
J_\da $ to $\Delta\vec A $ provides a measure of the polarisability of
individual Cooper pairs according to eq.~(\ref{polarisability}). In
the absence of a vector potential Cooper pairs are spherically
symmetric. A time-varying vector potential $\Delta\vec A $ acts like a
time-varying electric field coupling differently on the two components
of the Cooper pairs. This leads to a time-dependent polarisation of
the Cooper pairs and thus to a time-dependent spin current. For
$\omega >2\Delta $ the gauge field $\Delta \vec A $ has enough energy
to break Cooper pairs apart and the gas shows a response similar to
that of a gas of non-interacting particles.

\section{Conclusion}
\label{sec:conclusion}

We have investigated the response of a two-component Fermi gas to a
vector potential which couples differently to the two spin
species. The vector potential can be decomposed into a component
acting in the same way on the two spin species (the spin-symmetric
component) and a component acting in the opposite way on the two spin
species (the spin-asymmetric component). We have shown that in the
limit $\omega\to 0 $ the response to the spin-asymmetric component can
be neglected, where $\omega $ characterises the time dependence of the
vector potential. Thus only the response to the spin-symmetric
component remains from which the superfluid density can be deduced,
similar to what is discussed in
Ref. \onlinecite{measuring_sf_fraction_of_ultracold_atomic_gas-nigel_zoran10}.
We have also addressed the response of the  spin current to the
spin-asymmetric component for larger values of $\omega $. For $\omega \ll 2 \Delta $ the response can be described in terms of a polarisability of the superfluid, arising from the displacement of the average position of the spin up and spin down particles in the initially spherically symmetric Cooper pairs. This polarisability can be related to the strength with which the Cooper pairs in the Fermi gas are bound. For $\omega \gg 2 \Delta $ the response of the system is the same as for a non-interacting Fermi gas, due to the fact that the perturbation has enough energy to overcome the Cooper pair binding energy. The response of the system is maximal when $\omega = 2\Delta $ at which point the perturbation couples resonantly to the Cooper pair pair breaking process.
Our results show that probing ultracold Fermi gases using vector potentials is a fruitful way forward in the study of fermion many body physics.

\acknowledgments{We acknowledge 
useful discussions with Z. Hadzibabic, N.~Tammuz and E.~Eliel on the
experimental realisability. This work was supported by EPSRC Grant No. EP/F032773/1.}

\bibliographystyle{prsty}

\begin{thebibliography}{10}

\bibitem{chin}
C. Chin, R. Grimm, P. Julienne, and E. Tiesinga, Rev. Mod. Phys. {\bf 82},
  1225  (2010).

\bibitem{bloch07}
I. Bloch, J. Dalibard, and W. Zwerger, Rev. Mod. Phys. {\bf 80},  885  (2008).

\bibitem{fermionic_sf_with_imbalanced_populations-zwierlein-ketterle06}
M. Zwierlein, A. Schirotzek, C. Schunck, and W. Ketterle, Science {\bf 311},
  492  (2006).

\bibitem{pairing_and_phase_sep_in_pol_fg_partridge_hulet}
G. Partridge {\it et~al.}, Science {\bf 311},  503  (2006).

\bibitem{leggett_orig_bec_bcs-xover1980}
A. Leggett, {\em Modern Trends in the Theory of Condensed Matter} (Springer
  Verlag, Berlin, 1980).

\bibitem{Observation_of_a_Strongly_Interacting_Degenerate_Fermi_Gas_of_Atoms-o%
hara-thomas-science02}
K.~M. O'Hara {\it et~al.}, Science {\bf 298},  2179  (2002).

\bibitem{prod_of_long_lived_ultracold_li_molecules-cubizolles03}
J. Cubizolles {\it et~al.}, Phys. Rev. Lett. {\bf 91},  240401  (2003).

\bibitem{Creation_of_ultracold_molecules_from_a_Fermi_gas_of_atoms-regal_natur%
e03}
C. Regal, C. Ticknor, J. Bohn, and D. Jin, Nature {\bf 424},  47  (2003).

\bibitem{pure_gas_of_opt_trapped_mol-jochim03}
S. Jochim {\it et~al.}, Phys. Rev. Lett. {\bf 91},  240402  (2003).

\bibitem{obs_of_bec_of_molecules_zwierlein03}
M.~W. Zwierlein {\it et~al.}, Phys. Rev. Lett. {\bf 91},  250401  (2003).

\bibitem{chin04}
C. Chin {\it et~al.}, Science {\bf 305},  1128  (2004).

\bibitem{Evidence_for_Superfluidity_in_a_Resonantly_Interacting_Fermi_Gas-kina%
st-thomas04}
J. Kinast {\it et~al.}, Phys. Rev. Lett. {\bf 92},  150402  (2004).

\bibitem{Vortices_and_superfluidity_in_a_strongly_interacting_FG-zwierlein-nat%
ure05}
M. Zwierlein {\it et~al.}, Nature {\bf 435},  1047  (2005).

\bibitem{dalibardreview}
J. {Dalibard}, F. {Gerbier}, G. {Juzeli{\= u}nas}, and P. {{\"O}hberg},
  {Artificial gauge potentials for neutral atoms}, arXiv:1008.5378, 2010.

\bibitem{measuring_sf_fraction_of_ultracold_atomic_gas-nigel_zoran10}
N.~R. Cooper and Z. Hadzibabic, Phys. Rev. Lett. {\bf 104},  030401  (2010).

\bibitem{john}
S.~T. {John}, Z. {Hadzibabic}, and N.~R. {Cooper},  Phys. Rev. A {\bf 83},  023610  (2011).

\bibitem{pitaevskii_stringari_RMP_09}
S. Giorgini, L.~P. Pitaevskii, and S. Stringari, Rev. Mod. Phys. {\bf 80},
  1215  (2008).

\bibitem{jackson_2nd_ed}
J.~D. Jackson, {\em Classical Electrodynamics}, 2nd ed. (John Wiley \& Sons,
  New York, 1975).

\bibitem{forster75}
D. Forster, {\em Hydrodynamic Fluctuations, Broken Symmetry, and Correlation
  Functions} (W. A. Benjamin, Reading, Mass., 1975).

\bibitem{ben_and_atland06}
A. Altland and B. Simons, {\em Condensed Matter Field Theory} (Cambridge
  University Press, Cambridge, UK, 2006).

\bibitem{deGennes66}
P.~D. Gennes, {\em Superconductivity of metals and alloys} (W. A. Benjamin, New
  York, 1966).

\bibitem{gradshteyn_ryzhik_tables_of_integrals_series_and_products}
I.~S. Gradshteyn and I.~M. Ryzhik, {\em Table of Integrals, Series, and
  Products}, 5th ed. (Academic Press, London, 1994).

\bibitem{dipole_pol_of_trapped_sf_FG-recati_stringari06}
A. Recati, I. Carusotto, C. Lobo, and S. Stringari, Phys. Rev. Lett. {\bf 97},
  190403  (2006).

\bibitem{nozieres_Fermi_liquids2}
P. Nozi\`eres, {\em Theory of Interacting {F}ermi Systems} (Addison-Wesley,
  Reading, MA, 1997).

\bibitem{theory_of_quantum_liquids1_nozieres_1966}
P. Nozi\`{e}res and D. Pines, {\em The theory of quantum liquids} (Perseus
  Books, New York, 1966).

\bibitem{duine_viginale_spin_drag_FG2010}
R.~A. Duine, M. Polini, H.~T.~C. Stoof, and G. Vignale, Phys. Rev. Lett. {\bf
  104},  220403  (2010).

\bibitem{grimm_sf_density_2009}
S. Riedl {\it et~al.}, arXiv:0907.3814v2, 2009.

\end{thebibliography}

\end{document}